\documentclass[fleqn,usenatbib]{mnras}
\usepackage{newtxtext,newtxmath}

\usepackage[T1]{fontenc}
\usepackage{url}
\usepackage{hyperref}
\DeclareRobustCommand{\VAN}[3]{#2}
\let\VANthebibliography\thebibliography
\def\thebibliography{\DeclareRobustCommand{\VAN}[3]{##3}\VANthebibliography}
\usepackage{float}

\usepackage{graphicx}	
\usepackage{amsmath}	

\title[Algebraic Dipole Moment as Solar Cycle Predictor]
{Algebraic Quantification of the Contribution of Active Regions to the Sun's Dipole Moment: Applications to Century-scale Polar Field Estimates and Solar Cycle Forecasting}

\author[Pal et al.]{
Shaonwita Pal,$^{1,3}$\thanks{shao.physics@gmail.com}
and Dibyendu Nandy$^{1,2}$
\thanks{dnandi@iiserkol.ac.in}
\\
$^{1}$Center of Excellence in Space Sciences India, Indian Institute of Science Education and Research Kolkata, Mohanpur 741246, West Bengal, India\\
$^{2}$Department of Physical Sciences, Indian Institute of Science Education and Research Kolkata, Mohanpur 741246, West Bengal, India\\
$^{3}$Department of Technical Education, Training and Skill Development, Government of West Bengal, Newtown Rajarhat 700160, West Bengal, India\\
}

\date{Accepted XXX. Received YYY; in original form ZZZ}

\pubyear{2024}

\begin{document}
\label{firstpage}
\pagerange{\pageref{firstpage}--\pageref{lastpage}}
\maketitle

\begin{abstract}
The solar cycle is generated by a magnetohydrodynamic dynamo mechanism which involves the induction and recycling of the toroidal and poloidal components of the Sun's magnetic field. Recent observations indicate that the Babcock-Leighton mechanism -- mediated via the emergence and evolution of tilted bipolar active regions -- is the primary contributor to the Sun's large-scale dipolar field. Surface flux transport models and dynamo models have been employed to simulate this mechanism, which also allows for physics-based solar cycle forecasts. Recently, an alternative analytic method has been proposed to quantify the contribution of individual active regions to the Sun's dipole moment. Utilizing solar cycle observations spanning a century, here, we test the efficacy of this algebraic approach. Our results demonstrate that the algebraic quantification approach is reasonably successful in estimating dipole moments at solar minima over the past century -- providing a verification of the Babcock-Leighton mechanism as the primary contributor to the Sun's dipole field variations. We highlight that this algebraic methodology serves as an independent approach for estimating dipole moments at the minima of solar cycles, relying on characteristics of bipolar solar active regions. We also show how this method may be utilized for solar cycle predictions; our estimate of the Sun's dipole field at the end of cycle 24 using this approach indicates that solar cycle 25 would be a moderately weak cycle, ranging between solar cycle 20 and cycle 24. 

\end{abstract}




\section{Introduction} \label{sec:intro}
 
Our home star, the Sun, is a gigantic hot ball of plasma with inherent magnetic activity. Sunspots, strongly magnetized dark regions on the solar surface \citep{Hale1908}, serve as reliable indicators of this magnetic activity. Observations show that sunspot numbers undergo quasi-periodic variations following an approximately 11-year recurring cycle, known as the solar cycle \citep{Clark1978QJRAS, Schwabe1844AN, SCHATTEN2003, Hathaway2015}. Halfway through the solar cycle, the Sun's activity reaches its peak, or the solar maximum, with the highest number of sunspot emergences. During this maximum phase, its magnetic north and south poles flip, after which the Sun calms down until it reaches a solar minimum, indicating the beginning of a new sunspot cycle. During solar maximum, a more magnetically active Sun leads to frequent occurrences of magnetic outbursts and plasma outflows, such as solar flares and coronal mass ejections(CMEs). These phenomena significantly impact satellite operations, space-based technologies and the Earth's upper atmosphere \citep{SpcWeather, Solanki2002}. Therefore, understanding the dynamics of the solar cycle is crucial to be able to predict the Sun's magnetic activity and its consequences on space weather and planetary environments \citep{Petrovay2020b, Nandy2021, Bhowmik2023, Nandy2023}.

The magnetic cycle of the Sun can be explained through the Babcock-Leighton (BL) Solar Dynamo theory, which primarily establishes the interplay between the global poloidal field and the toroidal field in the presence of various plasma flows within the solar convection zone \citep{Wang1991, Leighton1964, Charbonneau2020}. During the initial phase of the solar cycle, the global magnetic field is primarily dominated by the poloidal field component. The Sun's differential rotation stretches this poloidal field in the longitudinal direction, leading to the formation of the toroidal field in the tachocline region \citep{Snodgrass1987SoPh}. Subsequently, these toroidal flux ropes are unstable within the convection zone, and due to magnetic buoyancy, they emerge on the solar surface as dark sunspots. Once the tilted bipolar active regions (BMRs) appear on the solar surface, their evolution and the regeneration of the toroidal field are primarily governed by the Babcock-Leighton (BL) mechanism \citep{Babcock1961}. The BL mechanism constitutes two processes: one is the annihilation of the leading polarities across two hemispheres, and the second one is the drift and diffusion of the following polarity towards the pole. These unipolar magnetic regions cancel the existing poloidal field at the pole and generate the poloidal field with opposite signs for the new solar cycle. Altogether, the BL-type Solar Dynamo model effectively captures the key aspects of the decay and dispersal of sunspots, polar field reversal, and the new polar field buildup \citep{CHARBONNEAU2007, Kitchatinov2011AstL, Cameron2017A&A, Bhowmik2018, Kumar2019A&A, Pal2023}.

It is well-established that during a solar activity minimum, the poloidal magnetic field, often referred to as polar field, and other polar field proxies (for example, axial dipole moment, A-t index, Geomagnetic aa-index etc.) strongly correlates with the amplitude of the succeeding cycle \citep{Schatten1978GeoRL, Yeates2008, Munoz2012}. Utilizing polar field proxies as a seed for predicting the amplitude of the next solar cycle is known as the `precursor method', which has evolved as one of the most successful techniques of sunspot cycle prediction \citep{Nandy2021, Petrovay2020b}. However, selecting an appropriate precursor for solar cycle forecasts relies on substantial physical insight and aids in accurate cycle predictions. In the context of the dynamo mechanism, the dipole moment (DM) closely relates to the poloidal field at the end of a solar cycle. Analysis of the observed photospheric magnetic field over the past four solar cycles suggests that the reversal of the dipole moment epoch aligns better with the cycle maximum than the average timing of polar field reversal \citep{Lisa2013, Iijima2017, Virtanen2019}. Moreover, the dipole moment contains information from the entire photosphere, mitigating the effects of a hemispherically asymmetric magnetic field distribution. Hence, during the solar minimum, the axial dipole moment component acts as a seed for the toroidal component of the next cycle \citep{Upton2018GeoRL, Charbonneau2020, Nandy2023arxiv}.

Predicting the dipole moment at the end of the solar cycle minimum is a feasible approach to estimate the strength of the next cycle. This task can be achieved through various methods, including observations and physics-based numerical models \citep{Lisa2013, Virtanen2019, Jaswal2023}. However, determining the dipole moment through magnetogram analysis is limited to a few past solar cycles and thus relies on physical models. One commonly used physics-based model for predicting the dipole moment is the Surface Flux Transport (SFT) model based on the BL mechanism \citep{Lisa2013, Bhowmik2018, Pal2023, Yeates2023SSRv}. However, calibrating such numerical models sometimes becomes challenging and time-consuming. What if we explore an alternative to numerical methods, moving away from complex computer-intensive modelling and adopting a simplified approach? 

The first attempt in this direction was made by \cite{Jiang2019ApJ, Petrovay2020a}. They introduced a mathematical framework aimed at calculating the distinct contributions of each emerging active region that collectively generate the ultimate global dipole moment during the cycle minimum. In their work, synthetic active region data was utilized to compute the ultimate dipole moment, and the results were compared with those derived from the 2 $\times$ 2D dynamo model \citep{Lemerle2017ApJ} simulations. Subsequently, \cite{Pal2023} adopted a similar approach to investigate the impact of anomalous active regions, specifically the combinations of synthetic Anti-Hale and Anti-Joy regions, on the solar cycle.

In this study, we employ the modified analytical approach to calculate the ultimate dipole moment at the end of a solar cycle, using the observational properties of bipolar active regions emerging throughout the declining phase of the sunspot cycle. Initially, we validate our method by estimating the dipole moment at the minima of solar cycles 14 to 23 and comparing it with observations. Subsequently, based on the algebraically derived dipole moment for solar cycle 24, we predict the peak amplitude of the ongoing solar cycle, \textit{i.e.} solar cycle 25, along with the associated uncertainties. We have also discussed the advantages, limitations and future possibilities of our methodology using observational insights.

\section{Methods} \label{sec:method_res}
Here, we discuss the method of quantifying the ultimate axial dipole moment of a solar cycle mathematically, which was first adopted by \cite{Petrovay2020a}. A spatially two-dimensional Surface Flux Transport model can be simplified to an azimuthally averaged 1D SFT model \citep{Petrovay2020a, Pal2023}. In this model, tilted sunspots transform into a bipolar flux ring with a finite latitudinal separation. Now, the `initial unsigned dipole moment' of any i$^{\mathrm{th}}$ active region can be expressed as,
\begin{eqnarray}
    \delta D_{1,i} = \frac{3}{4 \pi R_\odot^2}\,{\Phi}_i\,d_{\lambda_i} \cos \lambda_i.
    \label{eq1}
\end{eqnarray}

Here, $\lambda_i$ is the latitudinal position of the i$^{\mathrm{th}}$ sunspot and $\mathrm{R}_\odot$ is the solar radius. The term $d_{\lambda_i}$ denotes the latitudinal separation of the leading and following polarities of the i$^{\mathrm{th}}$ sunspot, $\Phi_i$ represents the magnetic flux content in the concerned sunspot. We take $d_{\lambda_i} = d_i \times \mathrm{sin}\alpha_i$, where $d_i$ represents the full angular polarity separation and $\alpha_i$ is the tilt angle of the sunspot relative to the east–west direction. In this study, we assume the tilt angle $\alpha_i$ to be proportional to the latitude expressed as $\alpha_i = 0.5 \times \lambda_i$ following \cite{Lemerle2015}. Additionally, we consider $d_i$ be proportional
to the radius of the sunspot \citep{Bhowmik2018}. Therefore, $d_{\lambda_i}$ is not constant for all sunspots and varies with the sizes and tilt angles of the active regions.

The evolution of the dipole moment involves additional physical factors that govern the regular dipole moment reversal and its accumulation. The dipole moment build-up may be influenced by the radial diffusion of the photospheric magnetic field. This radial diffusion term is expressed as $\mathrm{e}^{-t/\tau}$, where $\tau$ represents the exponential decay term. This expression indicates that the dipole moment gradually diminishes over time due to the radial outflows.

Another asymptotic dependency of the dipole moment is linked to the latitudinal position of sunspots. \cite{Jiang2014ApJ} demonstrated, through SFT simulations, that the amplitude of the dipole moment decreases with increasing latitude. This asymptotic dipole moment contribution factor, denoted as $f_\infty$, can be modeled as a Gaussian function of latitude:
\begin{eqnarray}
f_{\infty} = C\,e^{- \frac{\lambda^2}{ 2\lambda_R^2}}
\end{eqnarray}

This expression eventually takes care of the latitudinal quenching effect, indicating that low-latitude sunspots have a more significant impact on the ultimate dipole moment, while high-latitude active regions contribute less to its buildup. The dynamo effectivity range $\lambda_R$ and $C$ ($= A/\lambda_R$) can be taken as constant, which are determined by specific flux transport profiles, such as meridional circulation and turbulent diffusion assumed in different surface flux transport models \citep{Petrovay2020a, Nagy2020, Wang2021}. However, approximate analytic estimates of these constants are also possible based on first principles, observations and certain assumptions. Following \cite{Petrovay2020a}, $\lambda_R$ in the low-latitude limit can be expressed as,
\begin{eqnarray}
\lambda_R = \left[\sigma^2+\frac{\eta}{R_\odot^2 \Delta_v}\right]^{1/2}, \mathrm{where} \,\, \Delta_v =  \left(\frac{1}{R_\odot}\,\frac{dv}{d\lambda}\right) \vline_{\lambda=0}
\end{eqnarray}

Here, $\sigma$ represents the half-width of a Gaussian sunspot, $\Delta_v$ is the divergence of the meridional flow at the equator, and $\eta$ is the magnetic diffusivity. Following helioseismic observations, we consider a sinusoidal profile for the meridional circulation peaking at mid-latitude \citep{jiang2014, Petrovay2020a}. Thus, the meridional circulation velocity profile can be expressed as, $v = v_0\,\sin{2\lambda}$. The half-width of a sunspot is assumed to be proportional to its radius, which is determined from the mean of the distribution of the observed sunspot radii. We find $\sigma = 0.56^\circ$ (the maximum half-width can go up to $4^\circ$). The transport parameters, including diffusivity ($\eta$) and peak amplitude of meridional circulation ($v_0$), were constrained within observational ranges \citep{jiang2014}. Specifically, we take $\eta = 700 \,\mathrm{km}^2\mathrm{s}^{-1}$ and $v_0 = 10\, \mathrm{ms}^{-1}$ which results in $\lambda_R = 12.86^\circ$. We emphasize that changing $\sigma$ from $0.56^\circ$ to $4^\circ$ does not change $\lambda_R$ much ($\pm\,1^\circ$). We assume that these transport profiles act similarly for individual active regions and will not vary from one solar cycle to another, therefore, $\lambda_R$ is constant throughout our analysis. Additionally, we utilize the observational polar field data \citep{wso} to determine the constant $C$. We find $C$ = 5.48. The method of calibrating $C$ is described in the results section \ref{sec3.1}.

\begin{figure*}
    \centering
    \includegraphics[scale=0.36]{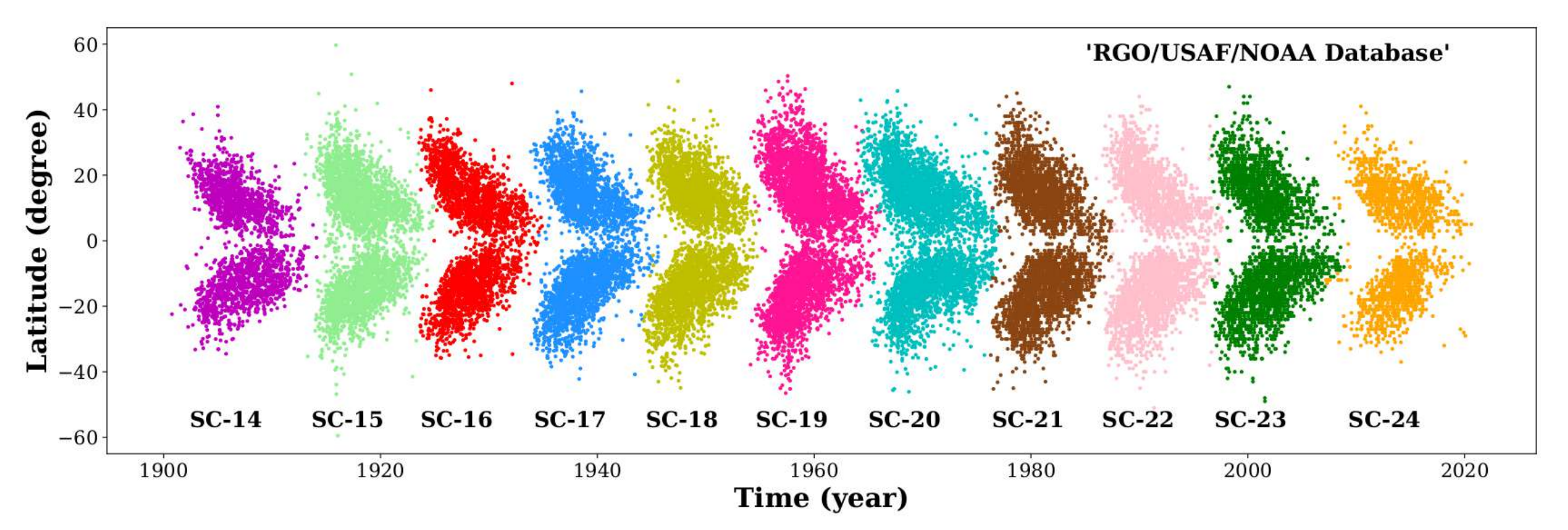}
    \caption{The butterfly diagram depicts the spatiotemporal changes spanning Solar Cycles 14 to 24. This figure illustrates the evolution of sunspots across the last 11 solar cycles, using time and latitudinal position data sourced from the \protect\cite{rgo}. Distinct colours are used to differentiate between the various solar cycles.}
    \label{fig:1}
\end{figure*}

Thus, if an i$^{\mathrm{th}}$ active region emerges at time $t_i$, then the ultimate dipole moment contribution from that active region at the end of cycle n at time $t_{n+1}$ becomes: 
\begin{eqnarray}
    \delta D_{\mathrm{U},i} = f_{\infty,i}\,\delta D_{1,i}\,\mathrm{e}^{\frac{(t_i - t_{n+1})}{\tau}}.  
    \label{eq11}
\end{eqnarray}

The initial dipole moment contribution from i$^{\mathrm{th}}$ sunspot ($\delta D_{1,i}$) will be positive or negative depending on the polarity of the sunspot that is closest to the equator. In general, one expects a positive contribution for Hale-Joy sunspots and a negative contribution for anomalous sunspots towards the ultimate dipole moment. In our recent study of anomalous active regions, we utilized this sign convention of dipole moment, revealing that the dipole moment decreases at solar minima when multiple anomalous active regions appear in a solar cycle \citep{Pal2023}. Here, for simplicity, we do not consider anomalous sunspots and confine ourselves to Hale-Joys active regions.

According to the BL mechanism, the sunspots of the current cycle decay and disperse due to plasma flows, cancelling the old cycle's dipole moment. Hence, the dipole moment at the end of a solar cycle is the combined result of the cancellation of the preceding cycle's dipole moment and the subsequent build-up of the new cycle's dipole moment. For brevity, in our study, we limit ourselves to active regions appearing after the reversal of the old cycle dipole moment to estimate the strength of the ultimate dipole moment of the cycle under consideration. In essence, therefore, we calculate the dipole moment contribution relative to its zero value (the latter happens approximately at solar maximum). The advantage of this methodology is that the calculation can be achieved even when the dipole moment at the end of the previous cycle is unknown.

Thus, the net contribution into the dipole moment at the end of n$^{\mathrm{th}}$ cycle from the active regions emerged during the n$^{\mathrm{th}}$ cycle can be expressed as the sum of the dipole moment contributions from individual sunspots that appeared after reversal time of that cycle. The analytic approach confirming this methodology is implemented as follows. The dipole moment at the end of cycle n is denoted by $\Delta \mathrm{DM}$ which is given by
\begin{eqnarray}
    \Delta \mathrm{DM} &=&  \sum_{i} \delta D_{\mathrm{U},i}\nonumber\\ &=& \sum_{i} f_{\infty,i}\,\delta D_{1,i}\,\mathrm{e}^{\frac{(t_i - t_{n+1})}{\tau}}\nonumber\\ &=&  \frac{3}{4 \pi R_\odot^2}\,\sum_{i} {\Phi}_i\,d_{{\lambda}_i} \cos {\lambda}_i\, C\mathrm{e}^{- \frac{\lambda_i^2}{2\lambda_R^2}}\nonumber\\ &=&  \frac{3}{4 \pi R_\odot^2}\,\sum_{i} {\Phi}_i\,d_i\sin {\alpha_i}\cos {\lambda}_i\, C\mathrm{e}^{- \frac{\lambda_i^2}{2\lambda_R^2}}\,.
    \label{eq3}
\end{eqnarray}

\noindent Here, `i' takes care of all active regions that emerge after the reversal of the dipole moment. We assume there are no radial outflows, \textit{i.e.} $\tau$ is infinity.

We compute the dipole moment at solar cycle minima for solar cycles 14 to cycle 24, utilizing the observed characteristics of the sunspots and the observed time reversal epoch of the dipole moment. We use the \cite{rgo} database to extract information on the latitudinal position and area of the bipolar active regions ranging from solar cycle 14 to cycle 24. In this study, we specifically focus on the statistics of active regions when they reach their maximum size. Figure~\ref{fig:1} illustrates the butterfly diagram, spanning the last century, with time and latitude information obtained from the RGO/USAF/NOAA database. We use the dipole moment reversal timing epoch from the WSO average polar field, available only for solar cycle 21 to cycle 24 \citep{wso}. For the rest of the solar cycles (\textit{i.e.} solar cycle 14 to cycle 20), we opt for the sunspot cycle peak time because the Sun's global dipole magnetic field generally flips its polarity around the maximum phase of the solar cycle. We extract the solar cycle maximum epoch from \cite{sidc} time series. Our analysis considers these dipole moment reversal timings as the standard reversal epoch.

\section{Results and Discussions}

\subsection{Dipole moment comparison for past solar cycles spanning a century.}
\label{sec3.1}
Utilizing the observational sunspot characteristics in the aforementioned analytical model, we estimate the century-scale calibrated global axial dipole moment at each solar minimum, spanning from solar cycle 14 to cycle 23. These algebraically derived dipole moments can be compared with observational dipole moment proxies, considering that the Sun's global magnetic field is primarily dipolar during the solar minimum. For this purpose, we use three observational time series: 1) Polar flux obtained from MWO polar faculae count \citep{Munoz2012}, 2) Makarov's A-t index \citep{Makarov2001SoPh}, and 3) WSO polar field \citep{wso}. We consider the average northern and southern hemispheric polar flux or polar field as a proxy for the dipole moment. 

Our mathematical approach focuses on determining the dipole moment's value at the end of the solar cycle rather than explaining its time evolution. Thus, we utilize the average polar field ($P_{av}$) obtained from the WSO Data Centre from 1976 onwards to calibrate the computed ultimate dipole moment for the last three solar cycles. We multiply all algebraic dipole moment (DM) values corresponding to each solar cycle minimum by the same constant factor $C$ and vary it until the DM versus $P_{av}$ is characterized by a line with a unit slope and zero intercept (i.e. we optimize the constant $C$ such that $P_{av} = C \times \mathrm{DM}$). This results in $C$ = 5.48. We select this particular constant as the calibration factor.

\begin{figure*}
    \includegraphics[scale=0.6]{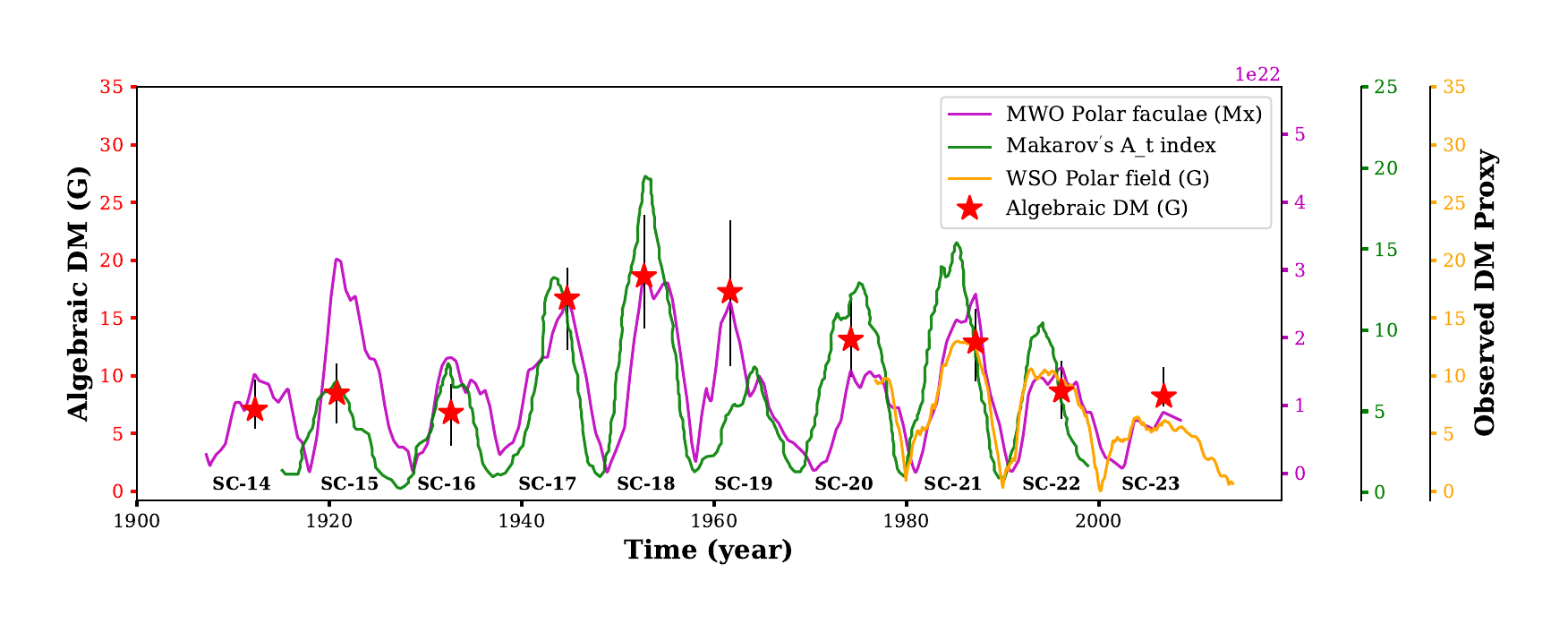}
    \caption{Time series of Dipole moment (DM) proxies ranging from solar cycle 14 to 24. In this representation, the MWO polar faculae data is depicted in magenta \protect\citep{Munoz2012}, Makarov's A-t index \protect\citep{Makarov2001SoPh} is shown in green, and the WSO polar field data \protect\citep{wso} is represented in orange. All polar field data is averaged from the northern and southern hemispheres to facilitate comparison with the dipole moment. Additionally, red stars indicate the algebraically computed ultimate dipole moment at the end of each cycle spanning solar cycle 14 to solar cycle 23.}
    \label{fig:2}
\end{figure*}

In Figure~\ref{fig:2}, red stars denote the estimated calibrated dipole moment compared with MWO polar flux, Makarov's A-t index and WSO polar field marked with magenta, green and orange. We calculate the potential error in the dipole moment computation by choosing the accurate dipole moment reversal time. The dipole moment reversal epoch may not always align with the solar cycle maxima; it can lead or lag the sunspot maximum epoch. If the dipole moment reversal timing lags the solar maximum epoch, then the total sunspots contribution towards the dipole moment will decrease, which in turn dampens the ultimate dipole moment. At the same time, the dipole moment will increase if the reversal time leads to the solar maximum epoch. Therefore, we assume that the dipole moment reversal time varies within a two-year interval around the standard reversal epoch, encompassing one year before and one year after the standard reversal epoch. Following that, we compute the contribution to the dipole moment at the endpoints of specified reversal time intervals. Based on this, we introduce an error bar on the derived dipole moment (see Figure~\ref{fig:2}). The algebraically derived dipole moments from solar cycle 14 to cycle 24 are tabulated in Table~\ref{table:1}.

\begin{figure*}
    \includegraphics[scale=0.5]{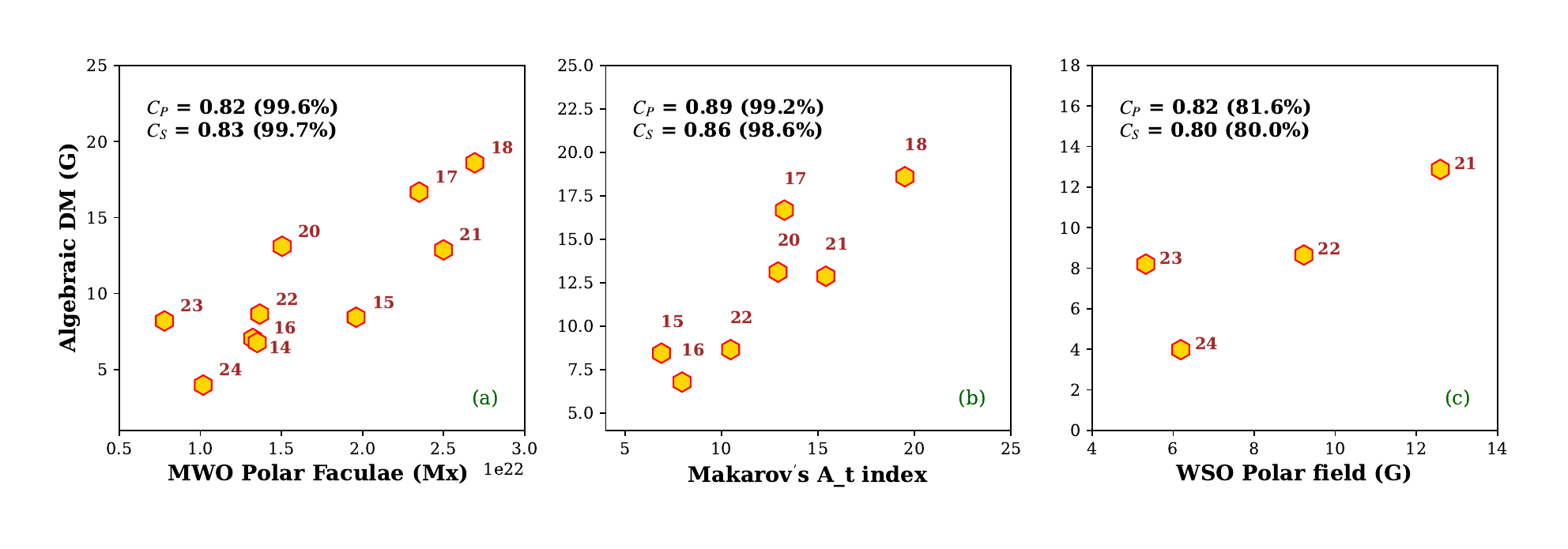}
    \caption{Statistical correlation analysis between observational dipole moment proxies and algebraic dipole moment. In Panel (a), (b), and (c), three different databases have been considered as the observed DM proxies: MWO polar faculae data \protect\citep{Munoz2012}, Makarov's A-t index \protect\citep{Makarov2001SoPh}, and WSO polar field \protect\citep{wso}. $C_P$ and $C_S$ denote Pearson and Spearman rank correlation coefficients, respectively.}
    \label{fig:3}
\end{figure*}

To check the efficacy of our methodology, we conduct a correlation analysis between the analytically calculated dipole moment and the observed polar field proxies at the solar minima. The results are depicted in Figure \ref{fig:3}. The correlation analysis demonstrates a reasonably good match between the algebraically derived dipole moment and the observed data, with one notable exception - solar cycle 19. The computed dipole moment for solar cycle 19, deviates significantly from the observational polar field proxies. Given this, we consider solar cycle 19 an outlier and exclude it from the correlation analysis.

After omitting solar cycle 19, we find a statistically significant correlation coefficient between the mathematically computed dipole moment and the observed polar field proxies, as mentioned in Figure \ref{fig:3}. In Section~\ref{sec:3.3}, we discuss potential factors contributing to our inability to retrieve the ultimate dipole moment of solar cycle 19. This study reveals that the analytically estimated dipole moment for the past ten solar cycles aligns with the observed polar flux. This alignment underscores the physics of decay and dispersal of sunspots, contributing to the ultimate build-up of the dipole moment from a mathematical perspective.

\subsection{Prediction of solar cycle 25.}
In this study, we find a significant deviation in the dipole moment of sunspot cycle 19. This deviation influences the peak amplitude of solar cycle 20, given the causal connection between the dipole moment at the solar minimum and the subsequent solar cycle strength. Hence, we exclude the dipole moment of solar cycle 19 and the peak amplitude of sunspot cycle 20 from our current analysis.

\begin{figure}
    \includegraphics[width=\columnwidth]{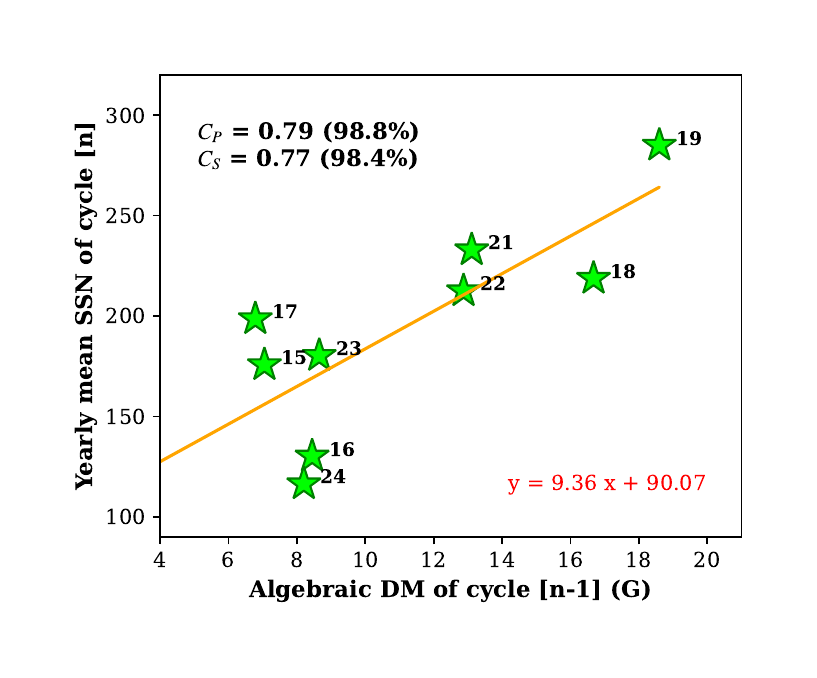}
    \caption{Statistical correlation analysis between analytically computed dipole moment of the cycle [n-1], denoted by `DM' and the yearly averaged sunspot number of the cycle [n] obtained from \protect\cite{sidc}, denoted by `N'. The scattered data points have been fitted with a linear regression model, visually represented by the orange line. The established relationship is expressed as follows: N = 9.36 $\times$ DM + 90.07.}
    \label{fig:4}
\end{figure}

We integrate the SIDC SILSO yearly averaged sunspot numbers dataset \citep{sidc} into our analysis to empirically predict ongoing solar cycle 25. First, we perform a correlation analysis between the analytically derived dipole moment at the end of a cycle [n-1] and the yearly mean sunspot number (SSN) of the consequent cycle [n]. The scatter plot in Figure \ref{fig:4} illustrates a strong positive correlation (with 99$\%$ confidence level) between these two quantities. This correlation suggests a potential empirical avenue for forecasting future cycles. Also, for the first time, we reconstruct the dipole moment spanning a century (from 1902 onwards) and utilize it for solar cycle prediction.

\begin{figure*}
    \includegraphics[scale=0.6]{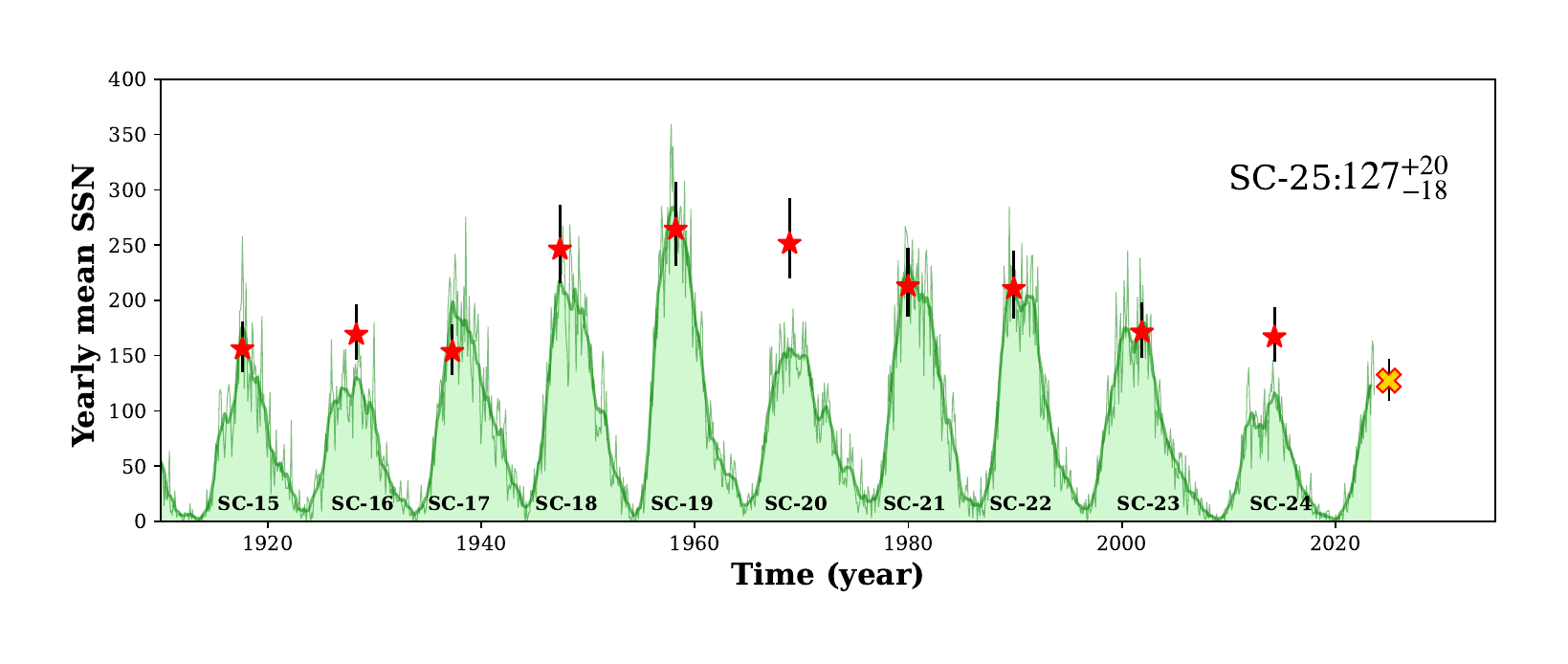}
    \caption{Solar Cycle 25 Prediction. The peak sunspot number for the last ten cycles is calculated and depicted as red stars. This is overlaid with the yearly averaged sunspot number time-series from \protect\cite{sidc}. Utilizing the ultimate dipole moment of cycle 24, the predicted amplitude of solar cycle 25 is $127_{-18}^{+20}$, denoted by the yellow cross.}
    \label{fig:5}
\end{figure*}

We fit this scatter plot in Figure \ref{fig:4} with a linear regression model and find a relationship between the analytically derived dipole moment (DM) and the sunspot number (N). The relationship is: 
N = 9.36 $\times$ DM + 90.07. Utilizing the ultimate analytic dipole moment at solar minimum (DM), we calculate the yearly average sunspot number (N) for solar cycle 14 to solar cycle 24. This method effectively reconstructs the past cycles, except solar cycle 20. Figure \ref{fig:5} shows that the empirically derived sunspot number at solar cycle maxima (red stars) is overplotted with the SIDC/SILSO sunspot numbers. This result is also tabulated in Table~\ref{table:1}. The deviation in cycle 20 is understandable, as our inability to accurately determine the cycle 19 dipole moment affects the subsequent cycle. 

Finally, by inputting the analytically computed ultimate dipole moment for solar cycle 24 into our fitted linear relationship, our prediction suggests that solar cycle 25 will be stronger than its predecessor, solar cycle 24. To be precise, we anticipate that solar cycle 25 will reach a yearly average peak sunspot number of 127, ranging between solar cycle 20 and cycle 24 (see Figure \ref{fig:5}). The reasonably good match of derived sunspot cycle maxima with the observed sunspot number over the last centuries also suggests that the dipole moment precursor is a promising candidate for solar cycle forecasts.

\begin{table}
\caption{Analytically derived dipole moment (DM) at the end of the solar cycle and predicted sunspot numbers, spanning last century (from solar cycle 14 to cycle 25).}

\label{T-complex}
\begin{center}
\begin{tabular}{ |p{2cm}|p{3cm}|p{2cm}|}
\hline
\textbf{Solar cycle $\#$} & \textbf{DM at solar minima (G)} & \textbf{Peak sunspot $\#$} \\
\hline
\hline
SC-14 & 7.05 [+2.6,-1.6] & \,\,\,\,\,\,\,\,\,\,\,\,\,\,\,\,- \\
\hline
SC-15 & 8.44 [+2.6,-2.6] & 156 [+25,-21] \\
\hline
SC-16  & 6.78 [+2.5,-2.8] & 169 [+27,-23] \\
\hline
SC-17 &  16.67 [+2.7,-4.4] & 153 [+24,-21] \\
\hline
SC-18 & 18.60 [+5.3,-4.6] & 246 [+40,-31] \\
\hline
SC-19 & 17.24 [+6.2,-6.4] & 264 [+43,-33] \\
\hline
SC-20 &  13.11 [+3.4,-3.2] & 251 [+41,-31] \\
\hline
SC-21 &  12.87 [+3.0,-3.3] & 213 [+35,-27] \\
\hline
SC-22 & 8.65 [+2.7,-2.4] & 211 [+34,-27] \\
\hline
SC-23 &  8.20 [+2.6,-0.9] & 171 [+27,-23] \\
\hline
SC-24 & 3.98 [+1.1,-1.5] & 169 [+27,-22] \\
\hline
SC-25 &  \,\,\,\,\,\,\,\,\,\,\,\,\,\,\,\, - & 127 [+20,-18] \\
\hline
\hline
\end{tabular}
\end{center}\label{table:1}
\end{table}

\subsection{Dependency of Algebraic Method on bipolar active region (BMR) characteristics.}\label{sec:3.3}
\begin{figure*}
    \includegraphics[scale=0.65]{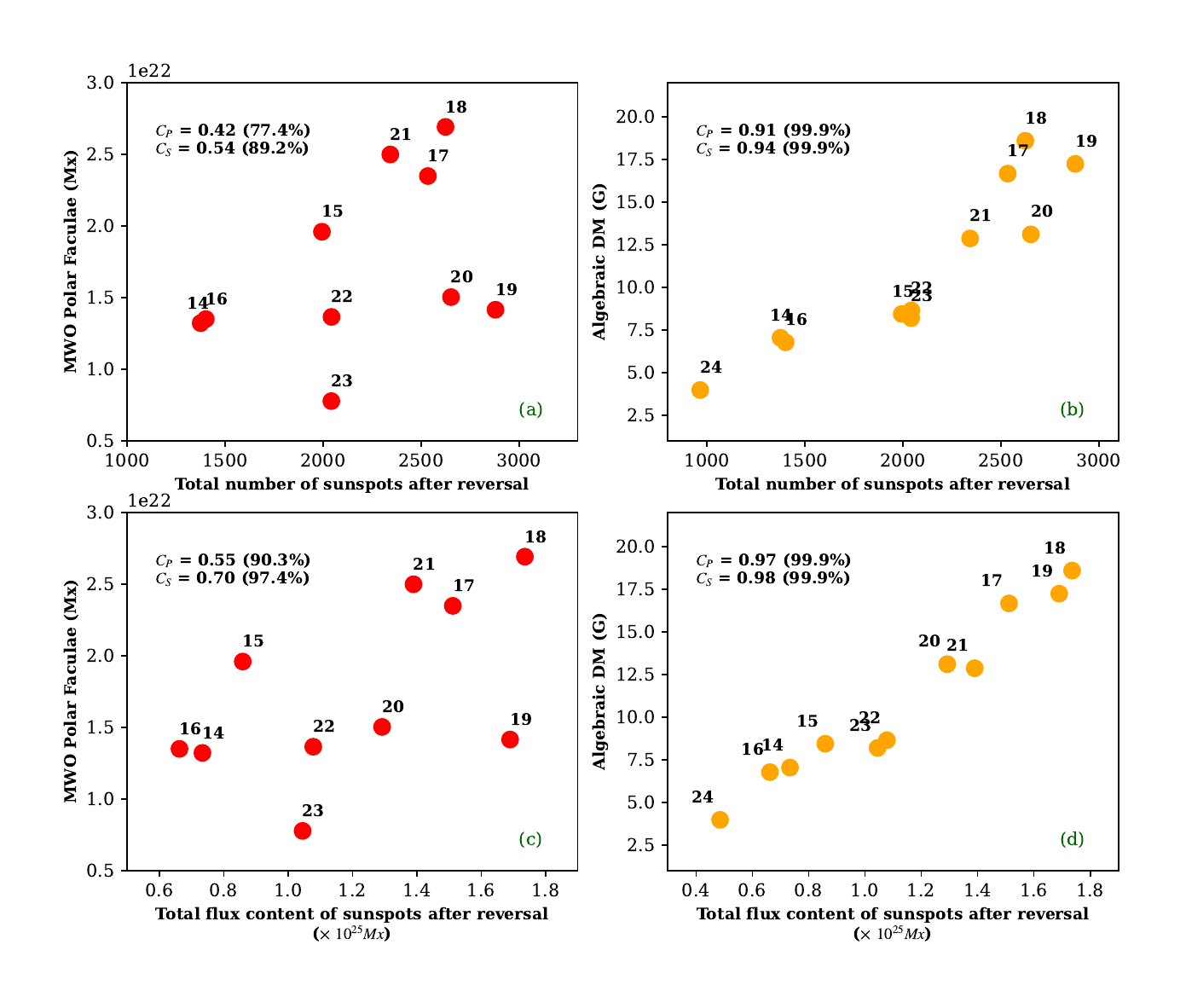}
    \caption{The correlation between the algebraic dipole moment (DM) and various properties of the Bipolar Magnetic Regions (BMRs).  In Panels (a) and (b), the correlation is observed between the total number of sunspots appearing after the dipole moment reversal and the observational average polar flux (Panel (a)) and algebraic dipole moment (Panel (b)). Similarly, panels (c) and (d) illustrate the correlation between the total flux content of all sunspots emerging after the reversal of the dipole moment and the observational average polar (Panel (c)) and algebraic dipole moment (Panel (d)).}
    \label{fig:6}
\end{figure*}

In this analytic model, the dipole moment at the solar cycle minima is sensitive to the quantity and flux content of active regions that emerge after the reversal of dipole moment polarity. To compare how this manifests in observation, we conduct a correlation analysis between the total number of active regions that appear after the time reversal and the 1) observed polar flux and 2) algebraic dipole moment at the end of the sunspot cycle. We find a high correlation between the total number of active regions and the ultimate algebraic dipole moments with a trend that is almost linear. This result is depicted in panel (b) of Figure \ref{fig:6}. Our finding indicates that as the number of sunspots increases, there is a monotonic rise in the ultimate dipole moment. However, this relationship is not as strong in observations which are indicative of a more non-linear relationship with the Spearman's rank correlation coefficient exceeding the Pearson's linear correlation coefficient (see Panel (a) in Figure \ref{fig:6}). Particularly during solar cycle 19, the highest number of sunspots appeared, but its dipole moment at the end of the cycle was notably small as illustrated in the same panel.

Similarly, we observe a strong correlation between the total flux content of the sunspots emerging after dipole moment reversal and the ultimate algebraic dipole moment estimated from our analytic method as depicted in Panel (d) of Figure \ref{fig:6}. However, the correlation with the observations depicted in Panel (c) is not as strong and is indicative of non-linearity.

The strong correlation between the number of sunspots and their magnetic flux content can significantly impact our algebraic calculations, occasionally yielding a dipole moment value much higher than expected. Due to this fact, we are unable to estimate the ultimate dipole moment of solar cycle 19 using algebraic techniques. Consequently, we cannot predict the yearly averaged sunspot number for solar cycle 20 very well.

Our analysis suggests that variations in BMR properties are not the sole factors contributing to the irregularities observed in the solar cycle. Other sources, such as nonlinear effects arising from fluctuations in meridional flows across cycles, plasma inflows, tilt quenching and anomalous active regions etc., which are not captured by the analytic method, may also play a significant role in accounting for the variability observed in the buildup of the ultimate dipole moment and, consequently, in modulating the solar cycle. Attempts to imbibe these nonlinear effects into the algebraic technique may lead to further improvements of dipole moment calculations and solar cycle predictions. These will be explored in the future.

\section{Conclusion}
In summary, we employ a simplified analytic technique, following the suggestion by \cite{Petrovay2020a} to calculate the Sun's dipole moment at the end of a solar cycle. We assess the effectiveness of algebraically derived dipole moments by comparing them with diverse observational data sets from solar cycle 14 to solar cycle 23. This analytic method reasonably estimates the observed dipole moment of solar cycles spanning a century. However, solar cycle 19 is an exception, being the strongest and most extreme cycle in the observation. 


Notably, for the first time, our method has allowed reconstruction of the dipole moment at solar minimum spanning a century -- allowing a robust test of its value as a precursor for predicting solar cycle amplitudes. We obtain a strong relationship between the dipole moment of the preceding cycle and the sunspot number of the subsequent solar cycle. Using this empirical relationship, we compute the yearly averaged sunspot number from solar cycle 14 to cycle 24. These estimates match the observations well. This is taken advantage of to predict the peak amplitude of sunspot cycle 25. We utilize the algebraically derived dipole moment of solar cycle 24 as a precursor to forecast the strength of solar cycle 25. The predicted amplitude is 128, with a range which places cycle 25 between sunspot cycle 20 and cycle 24.

Our work provides strong support to the idea that the emergence and evolution of tilted bipolar sunspot pairs are the primary contributors to the Sun's dipole field -- the so-called Babcock-Leighton mechanism. Surface flux transport models rely on this idea, and numerous dynamo models have been developed based on this idea, which reproduce diverse characteristics of the sunspot cycles. \citep{jiang2014, Cameron2015, Hazra2016ApJ, Hazra2019MNRAS, Saha2022, Dash2023, Karak2023, Pal2023, GHazra2023}

This analytic method for estimating the Sun's dipole moment relies on diverse properties of sunspots, including their latitudinal position, flux content, separation between two polarities and the total number of active regions that appear in a solar cycle. This method's dependency on BMR characteristics can sometimes lead to deviation from observations in the theoretical dipole moment calculation. For example, in our analysis, we observe a deviation in the dipole moment of solar cycle 19, which in turn impacts the reconstructed amplitude of solar cycle 20. This occurs because the direct dependency on the total number of sunspots and the flux content can weaken the accuracy of dipole moment estimates at the end of the cycle, which is not always seen in the observation. Nevertheless, such deviations are not always significant, and the dipole moment at solar minima derived from this analytic method closely matches the observational proxy based on polar flux. Notably, the sunspot amplitudes of past cycles empirically derived from the analytically estimated dipole moment match sunspot cycle amplitudes spanning the last century. Taken together, these corroborate our algebraic approach for dipole moment estimations. 

We emphasize that our methodology is an independent means to estimate the dipole moment at the minima of solar cycles based on characteristics of the sunspot time series. Therefore, this provides a straightforward theoretical tool to reconstruct the dipole moment at the minima of past sunspot cycles. We want to highlight that our method doesn't rely on solving numerical equations like those used in surface flux transport models.

There are certain shortcomings in such an approach, which relies on a 1D analytical model. For example, detailed studies of inter-active region interactions, exploration of non-axisymmetric phenomenon and influence of non-linearities can not be addressed. For these aspects, one must still rely on numerical simulations of spatially extended time-dependent magnetic field evolution models. However, when it comes to estimating the solar dipole moment and making predictions for the solar cycle, the algebraic method explored here appears to be a useful, independent tool.

\section{Acknowledgement}
This research was conducted at the Center of Excellence in Space Sciences India (CESSI), supported by IISER Kolkata and the Ministry of Education, Government of India. We acknowledge insightful discussions with Kristof Petrovay and members of the ISSI team entitled "What Determines The Dynamo Effectivity Of Solar Active Regions?" (supported by the International Space Science Institute in Bern, Switzerland). The authors also thank the referee for valuable suggestions.

\section{Data Availability}
We utilize the yearly and monthly averaged sunspot numbers from World Data Center SILSO, Royal Observatory of Belgium, Brussels \citep{sidc}. The dipole moment proxies are sourced from various datasets, including MWO polar faculae data \citep{Munoz2012}, Makarov's dipole-octupole index or A-t index \citep{Makarov2001SoPh}, and WSO polar field data \citep{wso}. Additionally, we extract Bipolar Magnetic Region (BMR) properties from the Royal Greenwich Observatory/USAF-NOAA active region database compiled by David H. Hathaway \citep{rgo}. The century-scale algebraically derived dipole moment data and solar cycle 25 prediction data will be made accessible upon reasonable requests. 


\bibliographystyle{mnras}
\bibliography{example} 











\bsp	
\label{lastpage}
\end{document}